\documentclass{PoS}

\usepackage{cite}

\def\beq{\begin{equation}}
\def\eeq{\end{equation}}
\def\beqa{\begin{eqnarray}}
\def\eeqa{\end{eqnarray}}

\title{Three-loop soft anomalous dimensions for top-quark production}

\ShortTitle{Three-loop soft anomalous dimensions for top-quark production}

\author{\speaker{Nikolaos Kidonakis}\thanks{This material is based upon work supported by the National Science Foundation under Grant No. PHY 1820795.}\\
Department of Physics, Kennesaw State University, USA\\
        E-mail: \email{nkidonak@kennesaw.edu}}

\abstract{I present results through three loops for soft anomalous dimensions that control soft-gluon emission in processes involving the top quark. In particular I present results for channels in single-top production and top-pair production as well as for processes with new physics, including $tZ$, $tZ'$, $t \gamma$, and $tH^-$ production. These calculations are ingredients to resummations at N$^3$LL accuracy and to derivations of N$^3$LO soft-gluon corrections.
}

\FullConference{
European Physical Society Conference on High Energy Physics - EPS-HEP2019 \\
                 10-17 July 2019\\
                 Ghent, Belgium}

\begin{document}

\section{Introduction}

Soft anomalous dimensions are fundamental field-theoretical functions that are essential in determining soft-gluon contributions to partonic scattering cross sections (for a review see Ref. \cite{NKtop}). Such cross sections factorize in Laplace or Mellin moment space into products of hard, soft, and jet functions.

The soft function, $S$, is in general a matrix in color space and it satisfies the renormalization group equation \cite{NKGS}
\beq
\left(\mu \frac{\partial}{\partial \mu}
+\beta(g_s)\frac{\partial}{\partial g_s}\right)\,S
=-\Gamma^\dagger_S \; S-S \; \Gamma_S
\eeq
with soft anomalous dimension $\Gamma_S$, also a matrix. The evolution of the soft function results in the exponentiation (resummation) of logarithms of the moment variable.
Resummation at NNLL accuracy requires two-loop soft anomalous dimensions while at N$^3$LL accuracy it requires three-loop soft anomalous dimensions.

In a top-quark production partonic process, $f_{1} +  f_{2} \rightarrow t + X$, 
we define a threshold variable $s_4$ that measures distance from partonic threshold. When the resummed cross section is inverted to momentum space, the soft-gluon corrections involve logarithms of $s_4$, but a prescription is needed for all-order results to deal with infrared divergences. Such resummed results are strongly prescription-dependent, and some prescriptions give poor results by underestimating the true size of the corrections. Alternatively, finite-order expansions can be performed with better control over subleading effects and no prescriptions. Expansions to second and third order (with matching with lower-order complete results) provide approximate NNLO (aNNLO) and N$^3$LO (aN$^3$LO) predictions which are state-of-the-art. As we approach partonic threshold, $s_4 \rightarrow 0$, there is diminishing energy left for additional gluon radiation, and the contributions from soft-gluon emission are dominant and they provide excellent predictions for the high-order corrections.

In the next section, we present results through three loops for the cusp anomalous dimension, which is the simplest soft anomalous dimension. We consider the case when both eikonal lines are massive and the case when one line is massive and one is massless. In Section 3, we present results through three loops for the soft anomalous dimensions in single-top production processes, in processes with a top quark in new physics models, and in top-antitop pair production.

\section{Three-loop cusp anomalous dimension}

\begin{figure}
\begin{center}
\includegraphics[width=0.3\textwidth]{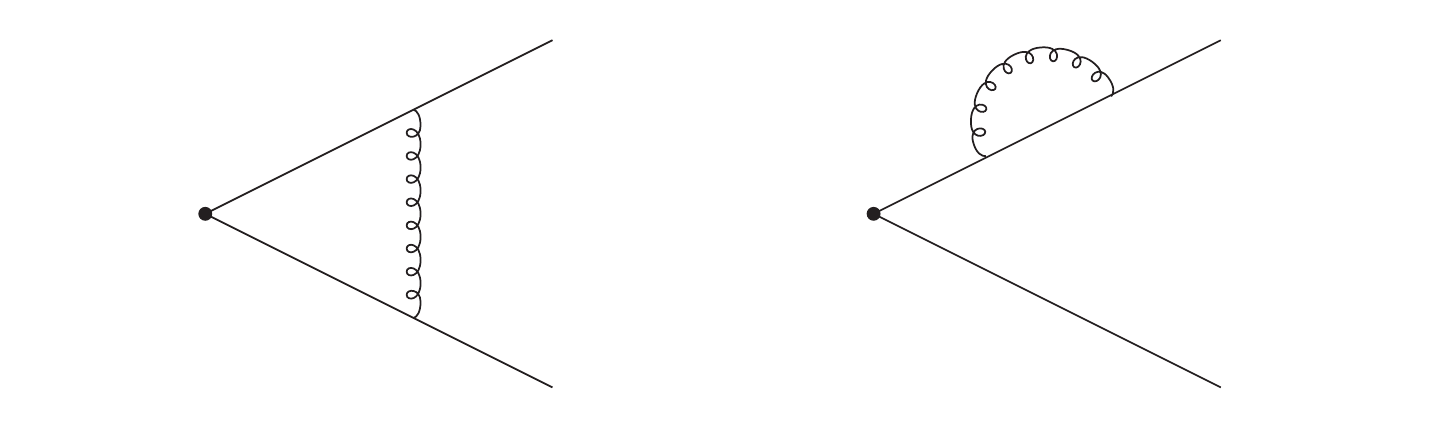}
\hspace{4mm}
\includegraphics[width=0.3\textwidth]{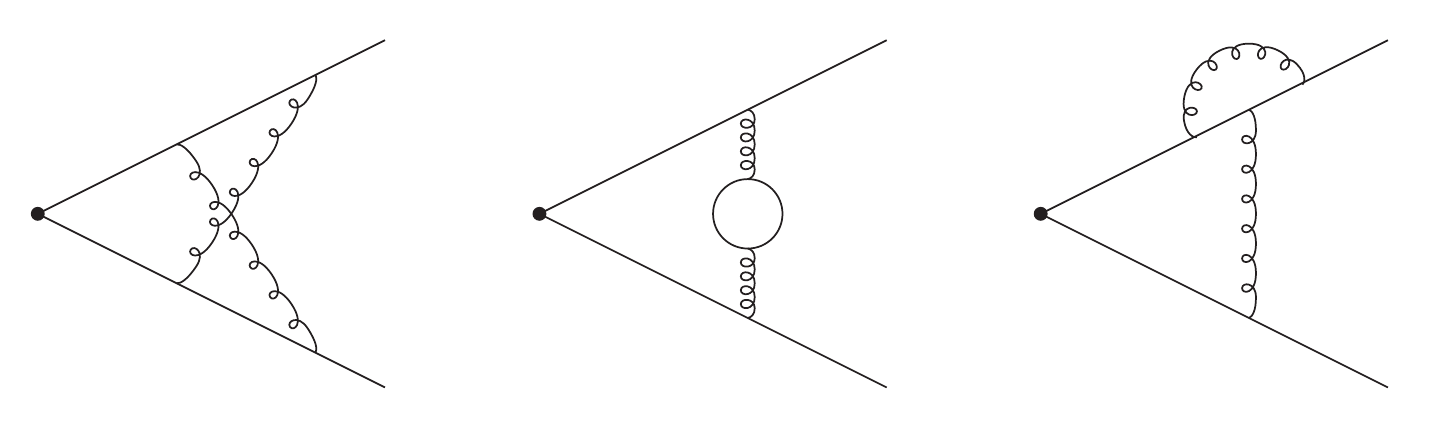}
\hspace{7mm}
\includegraphics[width=0.3\textwidth]{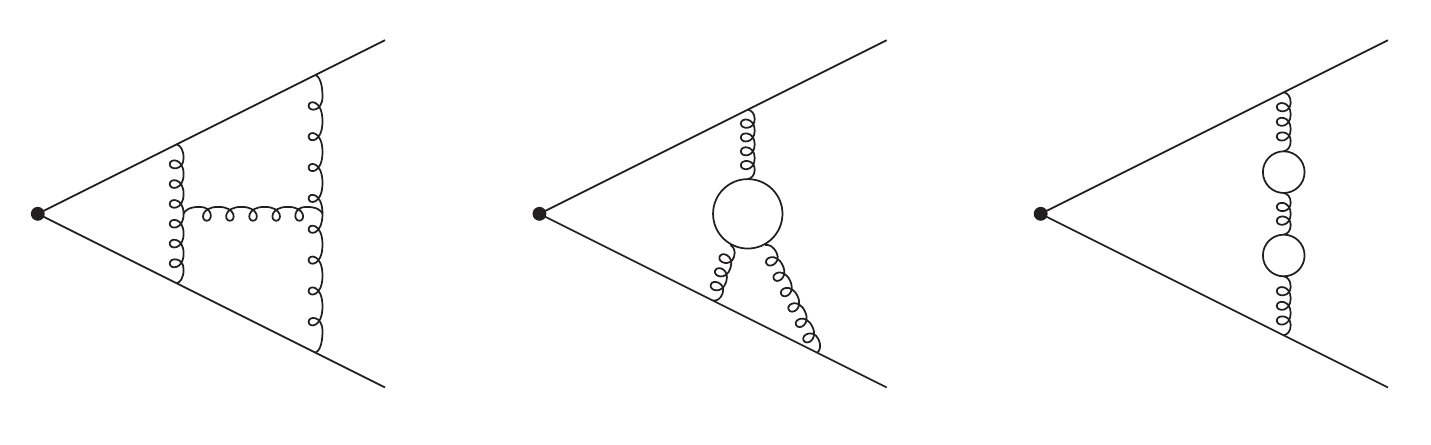}
\caption{Representative eikonal diagrams for the cusp anomalous dimension at one, two, and three loops.}
\label{cusp}
\end{center}
\end{figure}

A basic ingredient of soft anomalous dimensions for partonic processes is the cusp anomalous dimension \cite{KR,NK2loop,GHKM,NK3lcusp,NK3loop} (see Fig. \ref{cusp} for representative eikonal diagrams at one, two, and three loops).
The cusp angle is defined by $\theta=\cosh^{-1}(p_i\cdot p_j/\sqrt{p_i^2 p_j^2})$, in terms of the momenta of eikonal lines $i$ and $j$, and the perturbative expansion for the cusp anomalous dimension is $\Gamma_{\rm cusp}=\sum_{n=1}^{\infty} \left(\frac{\alpha_s}{\pi}\right)^n \Gamma^{(n)}_{\rm cusp}$. When the lines represent a top and an antitop, then the cusp anomalous dimension can be thought of as the soft anomalous dimension for $e^+ e^- \rightarrow t{\bar t}$. From the UV poles of the loop diagrams in dimensional regularization we calculate results at each order. 

At one loop,
\beq
\Gamma_{\rm cusp}^{(1)}=C_F (\theta \coth\theta-1) \, ,
\eeq
where $C_F=(N^2-1)/(2N)$, with $N$ the number of colors.
In terms of the heavy-quark speed $\beta=\tanh(\theta/2)$, we have
$\theta=\ln\left(\frac{1+\beta}{1-\beta}\right)$
and 
\beq
\Gamma_{\rm cusp}^{(1)}=C_F\left[-\frac{(1+\beta^2)}{2\beta}
\ln\frac{(1-\beta)}{(1+\beta)}-1\right] \, .
\eeq

At two loops \cite{NK2loop},
\beqa
\Gamma_{\rm cusp}^{(2)}&=&K^{'(2)} \, \Gamma_{\rm cusp}^{(1)}
+\frac{1}{2}C_F C_A \left\{1+\zeta_2+\theta^2 
-\coth\theta\left[\zeta_2\theta+\theta^2
+\frac{\theta^3}{3}+{\rm Li}_2\left(1-e^{-2\theta}\right)\right] \right. 
\nonumber \\ && \hspace{29mm} \left.
{}+\coth^2\theta\left[-\zeta_3+\zeta_2\theta+\frac{\theta^3}{3}
+\theta \, {\rm Li}_2\left(e^{-2\theta}\right)
+{\rm Li}_3\left(e^{-2\theta}\right)\right] \right\}
\eeqa
where $K^{'(2)}=K^{(2)}/C_F=C_A (67/36-\zeta_2/2)-5 n_f/18$, with $C_A=N$ and $n_f$ the number of light quark flavors.

At three loops \cite{GHKM,NK3lcusp},
\beq
\Gamma_{\rm cusp}^{(3)}= K^{'(3)} \Gamma_{\rm cusp}^{(1)}
+2 K^{'(2)} \left[\Gamma_{\rm cusp}^{(2)}-K^{'(2)}\Gamma_{\rm cusp}^{(1)}\right]
+C_{\rm cusp}^{(3)} 
\label{3lc}
\eeq
where $K^{'(3)}=K^{(3)}/C_F$ is a lengthy expression with color terms and constants (note that in general $K^{(n)}$ denotes the proportionality factor of the $n$-loop cusp anomalous dimension in the massless limit with $\theta$). The expression for $C_{\rm cusp}^{(3)}$ is very long (see \cite{NK3lcusp} for details). For top-quark production $n_f=5$, and a simple numerical expression is \cite{NK3lcusp}
\beq
\Gamma_{\rm cusp}^{(3)\, \rm approx}(\beta)=
0.09221 \, \beta^2+2.80322 \; \Gamma_{\rm cusp}^{(1)}(\beta) \, .
\label{3lca}
\eeq

For the case where one eikonal line is massive and one is massless we find simpler expressions for the cusp anomalous dimension, which we now denote as $\Gamma_c$ to distinguish it from the fully massive case. If eikonal line $i$ represents a massive quark and eikonal line $j$ a massless quark, then 
\beq
\Gamma_c^{(1)}=C_F \left[\ln\left(\frac{2 p_i \cdot p_j}{m_i \sqrt{s}}\right)
-\frac{1}{2}\right] \, ,
\eeq
\beq
\Gamma^{(2)}_c=K^{(2)} \left[\ln\left(\frac{2 p_i \cdot p_j}{m_i \sqrt{s}}\right
) -\frac{1}{2} \right]+\frac{1}{4} C_F C_A (1-\zeta_3) \, ,
\eeq
\beqa
\Gamma^{(3)}_c&=&K^{(3)} \left[\ln\left(\frac{2 p_i \cdot p_j}{m_i \sqrt{s}}\right)-\frac{1}{2}\right]+\frac{1}{2} K^{(2)} C_A (1-\zeta_3)
\nonumber \\ &&
{}+C_F C_A^2\left[-\frac{1}{4}+\frac{3}{8}\zeta_2-\frac{\zeta_3}{8}-\frac{3}{8}\zeta_2 \zeta_3+\frac{9}{16} \zeta_5\right] \, .
\eeqa

The cusp anomalous dimension is an essential component in calculations of soft anomalous dimension matrices for general partonic processes. In the fully massless case, two-loop soft anomalous dimension matrices are proportional to the one-loop quantity \cite{ADS}; however, when masses are present this is no longer the case. We also note that three-parton correlations with at least two massless lines vanish at any order due to constraints from scaling symmetry \cite{BN,GM}. However, four-parton correlations at three loops and beyond do not necessarily vanish even in the massless case \cite{ADG}. We will show explicit results at one, two, and three loops for various top-quark processes in the next section.

\section{Soft anomalous dimensions for top-quark processes}

We now present results for the soft anomalous dimensions through three loops for a variety of top-quark production processes.

\subsection{Single-top-quark production}

\begin{figure}
\begin{center}
\includegraphics[width=0.8\textwidth]{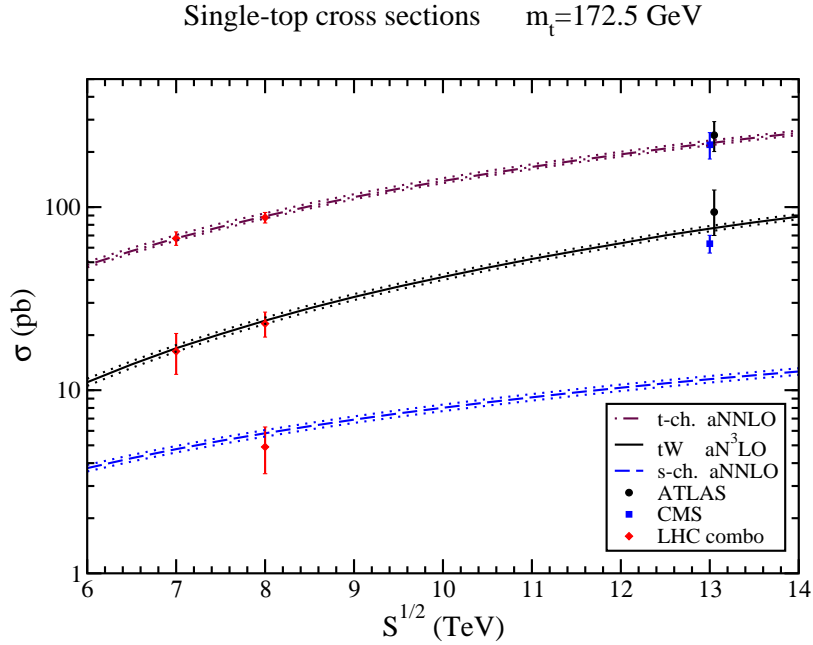}
\hspace{4mm}
\caption{Theoretical single-top cross sections for the $t$ and $s$ channels at aNNLO \cite{NKtop,NKsingletop} and for $tW$ production at aN$^3$LO \cite{NKtW}, using MMHT2014 NNLO pdf \cite{MMHT2014}, at LHC energies compared with LHC data \cite{ATLAS13tch,CMS13tch,LHC7and8singletop,ATLAS13tW,CMS13tW}.}
\label{singletop}
\end{center}
\end{figure}

We begin with single-top-quark production \cite{NKtop,NK3loop,NKst,NKsingletop,NKtW}. Before presenting the analytical form of the soft anomalous dimensions, we show in Fig. \ref{singletop} theoretical predictions at NNLL accuracy for the cross sections at LHC energies. The soft-gluon corrections are important and the theoretical predictions are in very good agreement with the data. 

\subsubsection{Single-top $t$-channel production}

The partonic processes in single-top $t$-channel production are 
$b(p_1)+q(p_2) \rightarrow t(p_3) +q'(p_4)$.
We define the usual partonic kinematical variables  
$s=(p_1+p_2)^2$, $t=(p_1-p_3)^2$, and $u=(p_2-p_3)^2$, and  
the threshold variable $s_4=s+t+u-m_t^2$.
We choose a singlet-octet $t$-channel color basis $c_1=\delta_{13} 
\delta_{24}$ and $c_2=T^c_{31} T^c_{42}$.
The soft anomalous dimension is a $2\times 2$ matrix.

At one loop, the four matrix elements are \cite{NKtop,NKst,NKsingletop}
\beqa
&&{\Gamma}_{S\, 11}^{t \, (1)}=
C_F \left[\ln\left(\frac{t(t-m_t^2)}{m_t s^{3/2}}\right)-\frac{1}{2}\right] \, , 
\quad
{\Gamma}_{S\, 12}^{t \, (1)}=\frac{C_F}{2N} \ln\left(\frac{u(u-m_t^2)}{s(s-m_t^2)}\right)
 \, , 
\quad
{\Gamma}_{S\, 21}^{t \, (1)}= \ln\left(\frac{u(u-m_t^2)}{s(s-m_t^2)}\right) \, ,
\nonumber \\ &&
{\Gamma}_{S\, 22}^{t \, (1)}= C_F \left[\ln\left(\frac{t(t-m_t^2)}{m_t s^{3/2}}\right)-\frac{1}{2}\right]-\frac{1}{N}\ln\left(\frac{u(u-m_t^2)}{s(s-m_t^2)}\right) 
+\frac{N}{2}\ln\left(\frac{u(u-m_t^2)}{t(t-m_t^2)}\right) \, .
\eeqa

At two loops, we find \cite{NK3loop,NKsingletop}
\beqa
&&\Gamma_{S\,11}^{t \, (2)}= K^{'(2)} \Gamma_{S\,11}^{t \, (1)}+\frac{1}{4} C_F C_A (1-\zeta_3)\, , 
\hspace{8mm}
\Gamma_{S\,12}^{t \, (2)}= K^{'(2)} \Gamma_{S\,12}^{t \, (1)} \, ,
\nonumber \\ &&
\Gamma_{S\,21}^{t \, (2)}= K^{'(2)} \Gamma_{S\,21}^{t \, (1)} \, , \hspace{8mm}
\Gamma_{S\,22}^{t \, (2)}= K^{'(2)} \Gamma_{S\,22}^{t \, (1)}+\frac{1}{4} C_F C_A (1-\zeta_3) \, .
\eeqa

At three loops, we only need the first element of the matrix for N$^3$LL resummation, and to calculate the N$^3$LO soft-gluon corrections. We have \cite{NK3loop}
\beq
\Gamma_{S\,11}^{t \, (3)}= K^{'(3)} \Gamma_{S\,11}^{t \, (1)}
+ \frac{1}{2} K^{(2)} C_A (1-\zeta_3) 
+C_F C_A^2\left[-\frac{1}{4}+\frac{3}{8}\zeta_2-\frac{\zeta_3}{8}
-\frac{3}{8}\zeta_2 \zeta_3+\frac{9}{16} \zeta_5\right] \, .
\eeq
Due to the simple structure of the leading-order hard matrix, the other three matrix elements of $\Gamma_S^t$ at three loops do not contribute to the N$^3$LO corrections. We can still provide expressions for those three matrix elements, up to four-parton correlations, as discussed in the context of $s$-channel production below.

\subsubsection{Single-top $s$-channel production}

The partonic processes in single-top $s$-channel production are 
$q(p_1)+{\bar q}'(p_2) \rightarrow t(p_3) +{\bar b}(p_4)$.
We again have a $2\times 2$ soft anomalous dimension matrix, and we choose a singlet-octet $s$-channel color basis, $c_1=\delta_{12} \delta_{34}$ and 
$c_2=T^c_{21} T^c_{34}$.

At one loop, the four matrix elements are \cite{NKtop,NKst,NKsingletop}
\beqa
&&\Gamma_{S\, 11}^{s \, (1)}=C_F \left[\ln\left(\frac{s-m_t^2}{m_t\sqrt{s}}\right)
-\frac{1}{2}\right] \, ,
\quad
\Gamma_{S\, 12}^{s \, (1)}=\frac{C_F}{2N} \ln\left(\frac{t(t-m_t^2)}{u(u-m_t^2)}\right) \, , 
\quad
\Gamma_{S\, 21}^{s \, (1)}= \ln\left(\frac{t(t-m_t^2)}{u(u-m_t^2)}\right) \, ,
\nonumber \\ &&
\Gamma_{S\, 22}^{s \, (1)}=C_F \left[\ln\left(\frac{s-m_t^2}{m_t \sqrt{s}}\right)-\frac{1}{2}\right]-\frac{1}{N}\ln\left(\frac{t(t-m_t^2)}{u(u-m_t^2)}\right)
+\frac{N}{2} \ln\left(\frac{t(t-m_t^2)}{s(s-m_t^2)}\right) \, .
\eeqa

At two loops, we have \cite{NK3loop,NKsingletop}
\beqa
&&\Gamma_{S\, 11}^{s\, (2)}=K^{'(2)} \Gamma_{S\,11}^{s \, (1)}+\frac{1}{4} C_F C_A (1-\zeta_3) \, , \hspace{8mm}
\Gamma_{S\,12}^{s \, (2)}=K^{'(2)} \Gamma_{S\,12}^{s \, (1)} \, ,
\nonumber \\ &&
\Gamma_{S\,21}^{s \, (2)}= K^{'(2)} \Gamma_{S\,21}^{s \, (1)} \, , \hspace{8mm}
\Gamma_{S\,22}^{s \, (2)}= K^{'(2)} \Gamma_{S\,22}^{s \, (1)}+\frac{1}{4} C_F C_A (1-\zeta_3) \, .
\label{sch2l}
\eeqa

At three loops, again we only need the first element of the matrix for N$^3$LL resummation and N$^3$LO corrections. We find \cite{NK3loop}
\beq
\Gamma_{S\, 11}^{s \, (3)}= K^{'(3)} \Gamma_{S\, 11}^{s \, (1)}
+\frac{1}{2} K^{(2)} C_A (1-\zeta_3) +C_F C_A^2\left[-\frac{1}{4}+\frac{3}{8}\zeta_2-\frac{\zeta_3}{8}
-\frac{3}{8}\zeta_2 \zeta_3+\frac{9}{16} \zeta_5\right] \, .
\label{sch3l}
\eeq
The other three matrix elements of $\Gamma_S^s$ at three loops are not fully known. However, we can see that the structure of the results will be similar to those at two loops.
The three-loop off-diagonal matrix elements will have a similar form as at two loops [replace all two-loop terms, denoted by superscript (2), in Eq. (\ref{sch2l}) by the corresponding three-loop terms, denoted by superscript (3)] while $\Gamma_{S\, 22}^{s \, (3)}$ should have the form of Eq. (\ref{sch3l}) [just replace the element subscript 11 with 22] up to four-parton correlations. The same also applies to the $t$-channel results as indicated previously.

\subsubsection{Associated $tW$ production}

In $tW$ production, the partonic process is
\beq
b(p_1)+g(p_2) \rightarrow t(p_3) +W^-(p_4) \, .
\eeq
In this case the soft anomalous dimension is a simple function (not a matrix).

At one loop, we have \cite{NKst,NKsingletop}
\beq
\Gamma_S^{tW \, (1)}=C_F \left[\ln\left(\frac{m_t^2-t}{m_t\sqrt{s}}\right)
-\frac{1}{2}\right] +\frac{C_A}{2} \ln\left(\frac{u-m_t^2}{t-m_t^2}\right)\, .
\eeq

At two loops, we have \cite{NKsingletop}
\beq
\Gamma_S^{tW (2)}=K^{'(2)} \Gamma_S^{tW \, (1)}
+\frac{1}{4}C_F C_A (1-\zeta_3) \, .
\eeq

At three loops, we have \cite{NK3loop}
\beq
\Gamma_S^{tW \, (3)}=K^{'(3)} \Gamma_S^{tW \, (1)}+\frac{1}{2} K^{(2)} C_A (1-\zeta_3)+C_F C_A^2\left[-\frac{1}{4}+\frac{3}{8}\zeta_2-\frac{\zeta_3}{8}
-\frac{3}{8}\zeta_2 \zeta_3+\frac{9}{16} \zeta_5\right] \, .
\eeq

\subsection{$tZ$, $tZ'$, $t \gamma$, and $tH^-$ production}

Top quarks can also be produced in association with electroweak and Higgs bosons in models of new physics. Such processes include the associated production of a top quark with a charged Higgs boson via $bg \rightarrow tH^-$ \cite{NKtH}; the associated production of a top quark with a $Z$ boson, $qg \rightarrow tZ$ \cite{NKtZ}, or with a photon, $qg \rightarrow t\gamma$ \cite{MFNK}, via anomalous couplings; and the associated production of a top quark with a $Z'$ boson either via anomalous couplings, $qg \rightarrow tZ'$, or via initial-state tops, $tg \rightarrow tZ'$ \cite{MGNK}.
The soft anomalous dimensions for all these processes are identical to the one for $bg \rightarrow tW^-$ that was presented in the previous subsection.

\subsection{Top-antitop pair production}

We continue with top-antitop pair production \cite{NKtop,NKGS,NKtt2l,NKtt}. The soft anomalous dimension is a $2\times 2$ matrix for the quark-initiated channel, $q{\bar q} \rightarrow t{\bar t}$, and a $3\times 3$ matrix for the gluon-initiated channel, $gg \rightarrow t{\bar t}$ \cite{NKGS,NKtt2l,FNPY}.

We begin with the soft anomalous dimension matrix for the $q{\bar q} \rightarrow t{\bar t}$ channel in a color tensor basis of $s$-channel singlet and octet exchange, $c_1 = \delta_{12}\delta_{34}$ and $c_2 =  T^c_{21} \, T^c_{34}$.

At one loop for $q{\bar q} \rightarrow t{\bar t}$ \cite{NKGS,NKtt2l}:
\beqa
\Gamma^{q{\bar q}\, (1)}_{S\, 11}&=&\Gamma_{\rm cusp}^{(1)} \, ,
\quad
\Gamma^{q{\bar q}\, (1)}_{S\, 12}=
\frac{C_F}{C_A} \ln\left(\frac{t-m_t^2}{u-m_t^2}\right) \, , 
\quad 
\Gamma^{q{\bar q}\, (1)}_{S\, 21}=
2\ln\left(\frac{t-m_t^2}{u-m_t^2}\right) \, ,
\nonumber \\ 
\Gamma^{q{\bar q}\, (1)}_{S\, 22}&=&\left(1-\frac{C_A}{2C_F}\right)
\Gamma_{\rm cusp}^{(1)} 
+4C_F \ln\left(\frac{t-m_t^2}{u-m_t^2}\right)
-\frac{C_A}{2}\left[1+\ln\left(\frac{s m_t^2 (t-m_t^2)^2}{(u-m_t^2)^4}\right)\right]. 
\eeqa

At two loops for $q{\bar q} \rightarrow t{\bar t}$ \cite{NKtop,NKtt2l}:
\beqa
\Gamma^{q{\bar q}\,(2)}_{S\, 11}&=&\Gamma_{\rm cusp}^{(2)} \, ,
\quad 
\Gamma^{q{\bar q}\,(2)}_{S\, 12}=
\left(K^{'(2)}-C_A N_S^{(2)}\right) \Gamma^{q{\bar q} \,(1)}_{S\, 12} \, ,
\quad
\Gamma^{q{\bar q} \,(2)}_{S\, 21}=
\left(K^{'(2)}+C_A N_S^{(2)}\right) \Gamma^{q{\bar q} \,(1)}_{S\, 21} \, ,
\nonumber \\
\Gamma^{q{\bar q} \,(2)}_{S\, 22}&=&
K^{'(2)} \Gamma^{q{\bar q} \,(1)}_{S\, 22}
+\left(1-\frac{C_A}{2C_F}\right)
\left(\Gamma_{\rm cusp}^{(2)}-K^{'(2)}\Gamma_{\rm cusp}^{(1)}\right) \, ,
\label{qq2l}
\eeqa
where 
\beq
N_S^{(2)}=\frac{\theta^2}{4}+\frac{1}{4} \coth\theta \left[\zeta_2-\theta^2-{\rm Li}_2\left(1-e^{-2\theta}\right)\right] \, .
\eeq

At three loops, it is evident that the diagonal elements of the three-loop matrix for $q{\bar q} \rightarrow t{\bar t}$ receive contributions from the three-loop massive cusp anomalous dimension, Eqs. (\ref{3lc}), (\ref{3lca}), but we do not yet have complete three-loop results. However, in analogy to our discussion for $s$-channel and $t$-channel single-top production, the structure of the results at three loops should be analogous to that at two loops [replace all two-loop terms, denoted by superscript (2), in Eq. (\ref{qq2l}) by the corresponding three-loop terms] up to four-parton correlations.

We continue with the soft anomalous dimension matrix for the $gg \rightarrow t{\bar t}$ channel in a color tensor basis $c_1=\delta_{12}\,\delta_{34}$, $c_2=d^{12c}\,T^c_{34}$, and $c_3=i f^{12c}\,T^c_{34}$. 

At one loop for $gg \rightarrow t{\bar t}$ \cite{NKGS,NKtt2l}:
\beqa
\Gamma^{gg\,(1)}_{S\, 11}&=& \Gamma_{\rm cusp}^{(1)}\, , \quad
\Gamma^{gg\,(1)}_{S\, 12}=0 \, , \quad \Gamma^{gg\,(1)}_{S\, 21}=0\, , \quad
\Gamma^{gg\,(1)}_{S\, 13}= \ln\left(\frac{t-m_t^2}{u-m_t^2}\right) \, , \quad
\Gamma^{gg\,(1)}_{S\, 31}= 2 \ln\left(\frac{t-m_t^2}{u-m_t^2}\right) \, ,
\nonumber \\
\Gamma^{gg\,(1)}_{S\, 22}&=& \left(1-\frac{C_A}{2C_F}\right)
\Gamma_{\rm cusp}^{(1)}
-\frac{C_A}{2}\left[1+\ln\left(\frac{s m_t^2}{(t-m_t^2)(u-m_t^2)}\right)\right] \, , 
\nonumber \\
\Gamma^{gg\,(1)}_{S\, 23}&=&\frac{C_A}{2} \ln\left(\frac{t-m_t^2}{u-m_t^2}\right) \, , \quad
\Gamma^{gg\,(1)}_{S\, 32}=\frac{(N^2-4)}{2N} \ln\left(\frac{t-m_t^2}{u-m_t^2}\right) \, , \quad \Gamma^{gg\,(1)}_{S\, 33}=\Gamma^{gg\,(1)}_{S\, 22} \, .
\eeqa

At two loops for $gg \rightarrow t{\bar t}$ \cite{NKtop,NKtt2l}:
\beqa
\Gamma^{gg\,(2)}_{S\, 11}&=& \Gamma_{\rm cusp}^{(2)} \, , \quad
\Gamma^{gg\,(2)}_{S\, 12}=0\, , \quad \Gamma^{gg\,(2)}_{S\, 21}=0\, , \quad 
\Gamma^{gg\,(2)}_{S\, 13}=\left(K^{'(2)}-C_A N_S^{(2)}\right) 
\Gamma^{gg \,(1)}_{S\, 13} \, , 
\nonumber \\ 
\Gamma^{gg\,(2)}_{S\, 31}&=&\left(K^{'(2)}+C_A N_S^{(2)}\right)  
\Gamma^{gg \,(1)}_{S\, 31} \, , \quad
\Gamma^{gg\,(2)}_{S\, 22}= K^{'(2)} \Gamma^{gg \,(1)}_{S\, 22}
+\left(1-\frac{C_A}{2C_F}\right) 
\left(\Gamma_{\rm cusp}^{(2)}-K^{'(2)}\Gamma_{\rm cusp}^{(1)}\right) \, ,
\nonumber \\
\Gamma^{gg\,(2)}_{S\, 23}&=& K^{'(2)} \Gamma^{gg \,(1)}_{S\, 23} \, , \quad
\Gamma^{gg\,(2)}_{S\, 32}= K^{'(2)} \Gamma^{gg \,(1)}_{S\, 32} \, , \quad
\Gamma^{gg\,(2)}_{S\, 33}=\Gamma^{gg\,(2)}_{S\, 22} \, .
\label{gg2l}
\eeqa

At three loops, again it is clear that the diagonal elements of the three-loop matrix for $gg \rightarrow t{\bar t}$ receive contributions from the three-loop massive cusp anomalous dimension, Eqs. (\ref{3lc}), (\ref{3lca}), but we do not yet have complete three-loop results. However, in analogy to our discussion for the  quark-initiated channel, the structure of the results at three loops should be analogous to that at two loops [replace all two-loop terms, denoted by superscript (2), in Eq. (\ref{gg2l}) by the corresponding three-loop terms] up to four-parton correlations.

\end{document}